# Revolutionizing Future Healthcare using Wireless on the Walls (WoW)

Jalil ur Rehman Kazim, Tie Jun Cui, Ahmed Zoha, Lianlin Li, Syed Aziz Shah, Akram Alomainy, Muhammad Ali Imran and Qammer H. Abbasi

*Abstract*— Following the standardization and deployment of fifth-generation (5G) network, researchers have shifted their focus to beyond 5G communication. Existing technologies have brought forth a plethora of applications that could not have been imagined in the past years. Beyond 5G will enable us to rethink the capability, it will offer in various sectors including agriculture, search and rescue and more specifically in the delivery of health care services. Unobtrusive and non-invasive measurements using radio frequency (RF) sensing, monitoring and control of wearable medical devices are the areas that would potentially benefit from beyond 5G. Applications such as RF sensing, device charging and remote patient monitoring will be a key challenge using millimetre (mmWave) communication. The mmWaves experience multi-path induced fading, where the rate of attenuation is larger as compared to the microwaves. Eventually, mmWave communication systems would require range extenders and guided surfaces. A proposed solution is the use of intelligent reflective surfaces, which will have the ability to manipulate electromagnetic (EM) signals. These intelligent surfaces mounted and/or coated on walls aka - 'Intelligent Walls' are planar and active surfaces, which will be a key element in beyond 5G and 6G communication. These intelligent walls equipped with machine learning algorithm and computation power would have the ability to manipulate EM waves and act as gateways in the heterogeneous network environment. The article presents the application and vision of intelligent walls for next-generation healthcare in the era of beyond 5G.

## I. INTRODUCTION

The The emerging paradigm of programmable and software-defined metasurfaces will enable unprecedented technical solutions to drive unobtrusive and non-invasive sensing that could revolutionize the health-care sector. The healthcare industry is one of the biggest and fastest-growing industry in the world. A report published by the Allied Market Research predicts that the healthcare market will reach $136.8 billion worldwide by 2021 [1]. Remote patient monitoring is expected to maintain its lead position with $72.7 billion until the same period. According to an estimate, there are 3.7 million medical devices in use that are connected to and/or monitor various parts of the human body to inform healthcare decisions. This has prompted researchers, academics and industry professionals to turn their attention to the introduction of the Internet of Medical things (IoMT). Over recent decades, there has been an enormous increase in the mobile and fixed wireless industry. In fact, mobiles have evolved from the analogue 1G voice system technology to the full-scale online and end-to-end digital 4G system (E2E). The B5G and future 6G wireless mobile communication systems will offer seamless access, enhanced Mobile Broadband (eMBB) with 1000x higher mobile data rates, Ultra-reliable and Low-Latency Communications (URLLC), i.e., 5x fewer delay optimised data [2-5]. The beyond 5G (B5G) will serve as an innovation engine across multiple sectors including but not limited to smart agriculture, pharmaceutical industry, intelligent health care and many more. The existing networks do not have the potential and resources to offer these services. In a wireless communication system, important network parameters could be optimized to achieve higher data rates and low latency. The crucial parameters are: (1) enhancing spectral efficiency, (2) deploying more base station, i.e., increasing network density, and (3) shifting to high bandwidth signals. In this context, the B5G physical layer would be embedded with three important breakthrough technologies. This includes Massive MIMO (Multiple-Input-Multiple-Output) antennas for spectral efficiency [6], Ultra-Dense Networks (UDNs) [7] and utilizing the mmWave spectrum, i.e. 30-300 GHz for communication.

In recent months some network operators have already started field tests and trials of 5G networks with Massive

[1]"This work was supported in parts by EPSRC EP/T021020/1 and EP/T021063/1." (*Corresponding author: Jalil ur Rehman kazim.*)

J. R. Kazim, A. Zoha, M. A. Imran, Q. H. Abbasi are with James Watt School of Engineering, University of Glasgow, Glasgow G12 8QQ, UK (e-mail: j.kazim.1@research.gla.ac.uk; ahmed.zoha@glasgow.ac.uk; Muhammad.Imran@glasgow.ac.uk; Qammer.Abbasi@glasgow.ac.uk).

T. J. Cui is with the State Key Laboratory of Millimetre Waves, Southeast University, Nanjing 210096, China (e-mail: tjcui@seu.edu.cn).

S. A. Shah is with Department of Computing and Mathematics, Manchester Metropolitan University, Manchester M15 6BH, UK (e-mail: S.Shah@mmu.ac.uk).

L. Li is with School of Electronic Engineering and Computer Sciences, Peking University, Beijing, 100871, China (e-mail: lianlin.li@pku.edu.cn).

A. Alomainy is with School of Electronic Engineering and Computer Science, Queens Mary University of London, London E1 4NS,UK (e-mail: a.alomainy@qmul.ac.uk).

MIMO technology operating in sub-6 GHz spectrum. One of the fundamental challenge faced by network operators is the usage of ultra-high frequency-mmWave signals. The mmWave frequencies are prone to path loss, attenuation and atmospheric absorption [8]. The communication of mmWave networks is inherently limited to point-to-point operation only. In a wireless network, the conventional approach is to control and optimise transmitter and/or the receiver side only. The wireless propagation channel has been considered as a black box due to its random nature. With the desire to control the propagation environment, researchers have proposed new concepts such as field programmable metasurface [9] and Intelligent Reflective Surface (IRS) [10-14]. With the ease of deployment and installation at both indoor and outdoor on any planar surface, they can be considered as 'Intelligent Walls' (IWs). With the ability to manipulate the electromagnetic (EM) waves they would be assistive technology in B5G and provide a vision about the application of IW assisted health care in B5G and 6G communication. The paper is organized as follows: In section II, a brief outline of the hardware architecture and mode of operation of IWs is introduced. Section III highlights the integration and control of machine learning (ML) with IW. In section IV, some important and potential use cases of IWs and their integration in various health care scenarios are briefly discussed. In the end, we provide potential future directions for IW in health care and future wireless communication.

## II. CONFLUENCE OF B5G AND INTELLIGENT WALLS

This integration of IWs will have an important role in the B5G communication ecosystem. The ability of mmWaves to travel long distances is inherently hampered because it diffracts less than the microwave signals.

Consequently, it suffers from signal blockage, atmospheric absorption and attenuation to a greater extent. As electronically steering guided systems, IWs allow these high-frequency signals to be steered in the desired direction. In contrast to a parabolic antenna and phased array systems, these are planar, RF chain free systems with some memory and intelligence. Primarily, the IW will be seen as a dependent entity embodied in future wireless systems.

## III. ARCHITECTURE AND OPERATION OF INTELLIGENT WALLS

A simplistic architecture of the IW is illustrated in Fig. 1. Hence, the IW can be divided into three sub-parts.

**EM layer:** It consists of a top layer which comprises simple metallic patches and active devices (e.g. PIN diodes) [9]. A conventional metal could be copper for mmWave signals or Gold/Graphene if Terahertz (THz) frequency is selected for operation [15]. The patches have sub-wavelength dimensions and inter-element spacing; giving them unique metasurface properties. A metasurface is an engineered surface that can be designed to perform or behave in a certain way. The reflecting elements on the surface are usually placed in a periodic manner. By controlling specific properties of the active devices to further engineer the phase response of the metasurface, the incoming signal direction can be steered and/or if required can be diffused [16, 17].

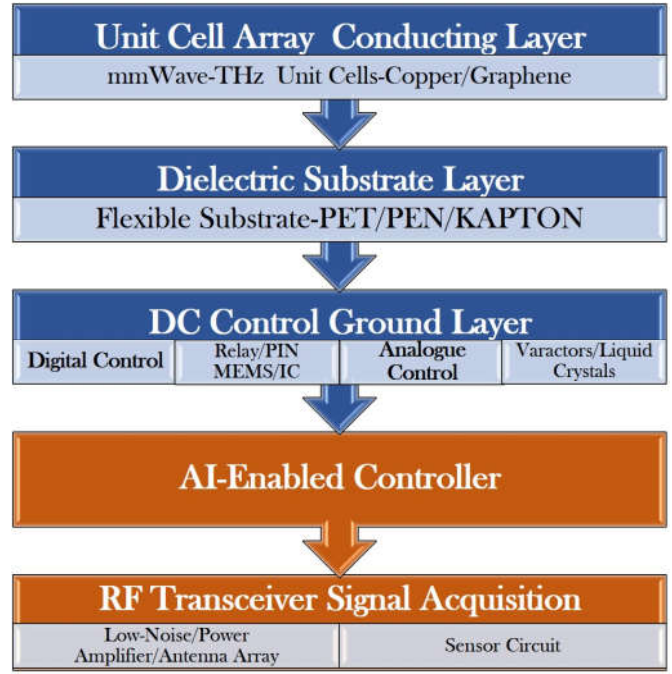

Fig. 1. A typical architecture of an intelligent reflective surface

**Dielectric layer:** Below the reflecting metal layer, this layer act as a dielectric spacer. The choice of the dielectric substrate is an important aspect of the IW design. An important design consideration is to have a thicker dielectric substrate with low dielectric constant. As a result, this would provide more element bandwidth [18]. The reflection efficiency may tend to decrease but can be improved by keeping the inter-element spacing below half-wavelength on the top metal layer [19]. Another obvious choice is to select flexible substrates with ease of deployment on any surface. The suitability of substrates is mainly dependent on the operating frequency. Each dielectric material provides a certain loss tangent at a particular frequency. Special attention is required if operating the IW at THz frequency. Conventional substrates at mmWave might not be a good choice in the ultra-high-frequency regime [20].

**Ground layer:** The third layer in the IW is the ground layer. The incident waves penetrating through the dielectric layer are reflected by the ground layer. At mmWave frequencies, this could be a simple copper metal.

**RF signal sensing:** A sub-component in the EM layer consists of a sensory circuit and an RF transceiver for RF signal acquisition and processing. Accordingly, the IW can be operated in two basic modes - Active mode and Passive mode.

**Passive mode**: This is the conventional operation of the IW. The ambient signals can be collected from the wireless channel and electronically steered or focused to a certain point in space [21].

**Active mode:** In this mode, the IW would send beacon signals to the base station with the instructions from the controller embedded with IW. The IW will operate in passive mode, only to be active with a request from the controller. Hence, this should not be confused with the normal transceiver operation where a continuous stream of information is sent over the channel and transceiver is always in the ON state [21].

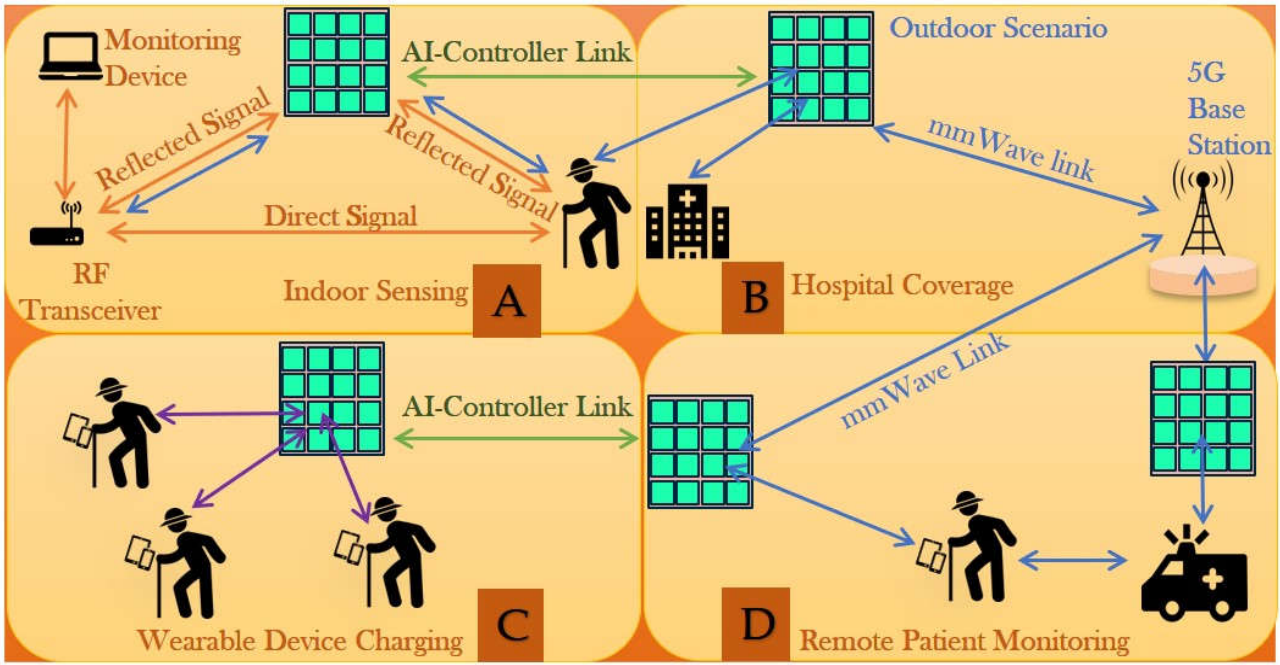

Fig. 2. Intelligent wall assisted different scenarios for health care use cases

**DC control layer:** The second sub-part that is the non-RF circuit is primarily responsible to control the EM layer. Two methods of control can be employed on this layer. An analogue control which can normally consist of varactor diodes [22], BST capacitors [23] and/or liquid crystal technology [24] having high-resolution tuning ability to control the phase of the unit cell in the EM layer. As the realization of continuous tuning and control is cumbersome in real-time, digital control method using PIN diodes [9, 25], MEM switches [26], and/or relays [27] could be adapted for discrete control of the EM layer. The concept of digitally controlled metasurfaces [9, 28] has been lately proposed and gained significant attention from the research community. This emerging concept has eventually converged the information science domain with EM domain giving rise to programmable metasurfaces [9].

**Controller Board:** The control algorithm is normally applied through the controller which could normally be a microcontroller (MCU) unit or a Field Programmable Gate Array (FPGA) device. The higher clock speed of an FPGA makes it suitable for applications where a faster switching operation is required.

## IV. INTELLIGENCE DRIVEN BY AI BRAIN

The functionality of reconfigurable metasurfaces, i.e., IWs, is mainly dependent on the sensing capability of the interface. Nonetheless, in typical practical scenarios, the optimization of many channel parameters based on sensed data, which is made available by sensors embedded into the IW, sets out requirements for large computing resource. To avoid continuous and iterative measurements and consume less power, Artificial Intelligence (AI) tools are envisioned to be embedded in the IWs to facilitate; by determining the best operation policy based on data-driven techniques [29]. The IW could be made adaptable, multi-functional and autonomous by incorporating AI into the system. Specifically, the incorporation of ML algorithms as a subfield of AI has revolutionized many realms of science and engineering and is foreseen to be enabled in IWs for a wide variety of applications. This stems from its capability to dynamically change the paradigm of data processing through the employment of algorithms that can learn from data and perform functionalities to complete complex tasks efficiently.

The concept has already been presented in some recent works [30-33]. A working prototype of a microwave reprogrammable digital metasurface [34] using ML algorithm is experimentally demonstrated. It is shown that the prototype device enabled with the ML algorithm can recognize the human movement in real-time. The proposed concept has paved the way for future AI-enabled software-controlled reflective surfaces for different health care services.

## V. INTELLIGENT WALLS FOR ASSISTIVE HEALTH CARE

The introduction of B5G communication will contribute to the commercial deployment of many healthcare technologies, including tele-diagnostic testing, telemedicine, ECG (Electrocardiogram) telediagnosis, ultrasonic evaluation, and online training of doctors. With in-hospitals scenario, B5G will open a wide range of opportunities and applications that require ultra-reliable low-latency links such as robotic surgery performed remotely and augmented reality aided surgery. These state-of-the-art applications require low latency and transfer of high-quality videos and very large image data. With a plethora of health care applications, we identify some important use cases with the potential to use these intelligent surfaces to enable these applications with B5G.

*A. Case I: Remote Monitoring/Telemedicine*

The challenge of an ageing population is pushing toward novel healthcare provisions that evolves from the traditional hospital-based system, where patients are treated or monitored for severe health conditions in a controlled environment, to a more person-centric approach. These patients can be observed in their homes remotely via 5G technology by identifying critical events such as freezing of gait and wandering behaviour. A home environment may typically block the incoming mmWave signals due to obstructions such as doors, windows and walls. The use of IW at designated places within the house would steer the beams to allow signals to reach the patient monitoring devices. These monitoring devices would typically relay back the patient data through these smart walls linking to a hospital network server. A typical scenario is depicted in Fig. 2, scenario A.

*B. Case II: Indoor Coverage and Localisation In Hospitals*

The B5G deployment may get complicated considering the resources and infrastructure requirements. Future wireless systems will use mmWaves which are located at the very high end of the spectrum, indoor communication and localisation will pose another challenge. Hospital buildings have metallic windows and metal frame-walls which will eventually block or either scatter the mmWave signals coming from 5G base stations. Due to planar nature and ease of deployment, these intelligent walls can be placed in corridors and around the corners within the hospital building for efficient indoor coverage. Localization involving mmWave frequencies will improve the accuracy from 10 m to below 1 m area [35], but the signals may not be able to propagate indoors efficiently. As IWs can create anomalous reflections, they will be able to assist in localization, mapping and tracking; by manipulating the signals to the desired locations [21, 36].

*C. Case III: Charging Wearable Devices*

The IoMT can provide a better way to care for our patients and has a tremendous potential to help deal with the rising costs of care. The IoMT refers to a system of interconnected medical devices and applications that collect data and forward it to healthcare IT systems. These medical devices are typically referred to as wearable health care devices. One problem associated with these devices is that they need to be frequently charged; to keep them active and send patient data to the monitoring systems. An important application of the IW will be to provide a focused beam of energy to wirelessly charge wearable devices [21, 37]. The sensing circuit associated with the IW would listen to the beacon signal of the wearable device and focus the ambient RF energy towards the device when a specific battery threshold is reached.

*D. Case IV: Real-Time Feedback for Preventive Care Applications*

A more specific scenario in assistive health care includes non-invasive sensing of patients [21, 38]. The mmWave signals could be used as an illuminator of opportunity where an ambient signal from 5G base station will be cross-correlated with echoes reflected from the subjects in the area under test, to extract information on their movements. With the help of ambient 5G signals, they can obtain full-scene images with high resolution and recognise human-body postures and vital signs with high accuracy in a smart, real-time and inexpensive way [21]. The processing can be performed at the AI-enabled controller connected to the IW. The IW will act as a primary surface mounted indoors. In a situation where the patient has a symptom of a heart attack or an elderly patient falls over, the AI-enabled controller would instruct a secondary IW to redirect the ambient mmWave signal to the dedicated emergency response radio link. The IW could be mounted outdoor communicating with a 5G base station. It is important to mention that the AI controller could be connected to multiple IWs with different operational requirements. The application is demonstrated in Fig. 2, scenario A and D.

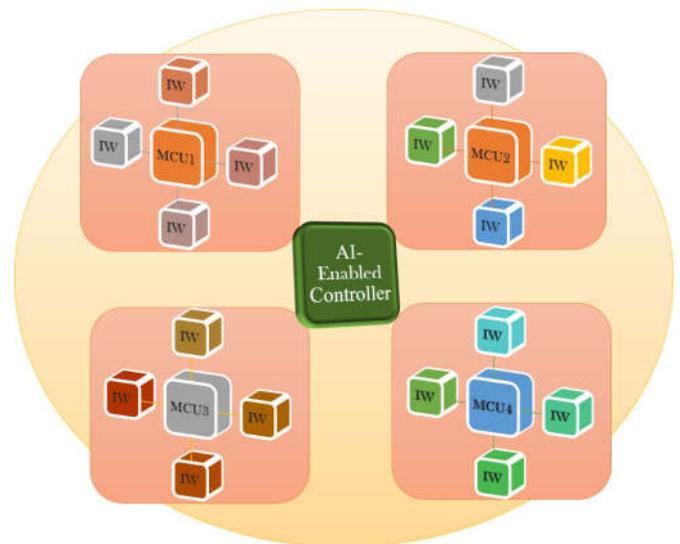

Fig. 3. Envisioned cluster network of Internet of Intelligent Walls enabled by AI controller

## VI. FUTURE DIRECTION

The Internet of Humans (IoH) is a novel concept in health care, enabling elderly and patients to be connected, monitored and recorded via the internet. People can keep track of their health and fitness with wireless wearable technologies. It is therefore important to foresee how IWs could overcome new technological barriers in the coming future.

**mmWave-THz Imaging with IW:** The ability of IW as an imaging system has already been demonstrated using WiFi signals [21]. This ability can be extended to operate in the mmWave and THz regimes as well. As the technology nurtures, we will expect to see miniaturized IW system used for diagnostics and imaging of patients.

**Internet of Intelligent Walls (IoIW):** The IW could be operated in a distributed network of IoIW. Analogous to a sensor network, the IW could be made to function as a plug and play device. A network of IWs connected with an MCU could be operated by AI-enabled controller link. The AI-enabled controller could be connected to several MCUs forming a cluster network of IWs. This could be interconnectivity of different hospitals and emergency services using IW.

**Integration of IW in BAN:** The IEEE 802.15.6 –Body Area Networks (BAN) protocol is specifically designed for on-body communication. The wearable devices using this protocol operate in a strict battery saving mode and hence the signal connectivity in the practical scenario might be very low. The connectivity problem in low power mode could be solved with the integration of IW within BAN. The IW could collect weak signals from these devices and hence interlink with another wearable device in the vicinity. In this regard, future BANs could be depicted as shown in Fig. 3.

**Self-Adaptive IWs**: Recently a self-adaptively smart metasurface has been proposed, in which a sensor or multiple sensors are integrated into the metasurface [39]. When the sensors detect different signals of the environment, the metasurface will automatically switch its functionalities without human operations. Based on this idea, a self-adaptive IW can be generated for smart home and health care, which could automatically operate in different modes under various environment conditions (e.g. daytime or night; temperature; humidity).

## VII. CONCLUSION

In this paper, we presented an overview of IWs and its potential to be integrated with future health care systems using B5G communication. The adaptation of mmWave spectrum for future wireless communication will provide users with enhanced data throughput, ultra-low latency and reliable communication. The requirement of broad coverage of mmWaves can be reached by integrating IWs in the wireless network. The IWs could smartly manipulate the signals and provide a full-duplex, interference-free communication link between the user and the base station. In the health care sector, the IWs can be seen charging a wearable device, monitoring patients in a non-invasive way and trigger to an emergency response unit by effective vital sign monitoring with AI-enabled technology. Through the presented discussion we can envision the integration and adaption of IW enabled communication with AI-enabled algorithms for future health care systems.